\documentclass[12pt]{JHEP3}
\usepackage{amssymb}
\usepackage{slashed}
\usepackage{bbold}
\usepackage{graphicx}

\title{Supergravity solutions for D-branes in Hpp-wave backgrounds}
\preprint{\hepth{0205106}\\DAMTP-02-51\\IC/2002/27}
\author{P. Bain\\DAMTP, University of Cambridge, \\ 
Centre for
Mathematical Sciences, Wilberforce Road, \\ 
Cambridge CB3 0WA, UK\\
{\tt E-mail: P.A.Bain@damtp.cam.ac.uk}} 
\author{P. Meessen\\
International School for Advanced Studies, \\ 
Via Beirut 2-4,\\ 34014
Trieste, ITALY\\
{\tt E-mail: meessen@sissa.it}} 
\author{M. Zamaklar\\
The Abdus Salam ICTP, \\ 
Strada Costiera 11, \\ 
34014 Trieste,
ITALY\\
{\tt E-mail: mzama@ictp.trieste.it}}

\abstract{We derive two families of supergravity solutions describing
D-branes in the maximally supersymmetric Hpp-wave background.  The
first family of solutions corresponds to quarter-BPS D-branes. These
solutions are delocalised along certain directions transverse to the
pp-wave. The second family corresponds to the non-supersymmetric
D-branes. These solutions are \emph{fully localised}. A peculiar
feature of the nonsupersymmetric solutions is that gravity becomes
repulsive close to the core of the D-brane.  Both families preserve
the amount of supersymmetry predicted by the D-brane probe/CFT
analysis.  All solutions are written in Brinkman coordinates.  To
construct these kind of solutions it is crucial to identify the
coordinates in which the ansatz looks the simplest.  We argue that the
natural coordinates to get the supergravity description of the
half-BPS branes are the Rosen coordinates.}

\begin{document}

\maketitle

\section{Introduction and Summary}

Recent developments of the AdS/CFT correspondence \cite{bmn} have
involved the study of a particular class of supergravity
pp-waves~\cite{gueven,blau}. These geometries naturally arise by
taking the so-called Penrose limit of various near horizon regions of
\mbox{D-branes}.  The Penrose limit~\cite{penrose, gueven} is a
general procedure in which, by a suitable rescaling of coordinates and
parameters characterising a (super)gravity solution, one focuses on
the region close to an arbitrary null geodesic. The scaling is
performed in such a way that the zoomed part of the space still solves
the equations of motion. In~\cite{penrose} Penrose proved that this
procedure leads, for \emph{any} gravity background, to a gravitational
pp-wave. The extension of his arguments to supergravity solutions,
which include additional higher rank forms, leads to generalisations
of gravitational pp-waves, which are, in addition to a nontrivial
metric, characterised by fluxes of NSNS or/and RR fields.

Waves which arise as Penrose limits of supergravity solutions for
various branes are also interesting from the pure supergravity
point-of-view since they usually preserve a fraction of supersymmetry
larger than one half \cite{clp,hull}.  In particular, the Hpp-wave
which will be considered in this paper is a maximally supersymmetric
vacuum of the type IIB string theory.

The main obstacle in proving the  conjectured AdS/CFT duality 
\cite{maldacena1,witten1,gkp}\footnote{For a more complete list of
references in  this area see \cite{maldacena2}.} 
 lies in the difficulty to quantise strings in AdS spaces.
The beauty of the proposal of \cite{bmn,horat} is that it suggests a
relatively simple class of supergravity backgrounds where one can
explicitly quantise strings \cite{metsaev, metsaev1} and ``identify''
a subsector of the corresponding gauge theory duals.  In \cite{blau1}
it was shown that taking the Penrose limit of the neighborhood of the
null geodesics along the 5-sphere in $AdS_5 \times S^5$ space leads
to a maximally supersymmetric gravitational pp-wave with a non
vanishing RR flux~(\ref{KG_10})~\cite{blau}. Furthermore, it was
argued that string excitations in this background correspond to the
subsector of the Yang-Mills theory characterised by large
R-charges. These states are dual to the string excitations in the
initial $AdS$ space that carry large angular momentum
($J\leftrightarrow R$) in one of the directions of the $S^5$.

In the AdS/CFT correspondence, D-branes usually correspond to
nonperturbative 
objects on the gauge side or defects on which a lower dimensional
conformal field theory lives.  In order to study the field theory
dynamics in the presence of these objects using the duality, it is
useful to have access to the supergravity solutions for D-branes in
$AdS$ spaces. Unfortunately, not much is known explicitly about these
solutions.  There are two methods that one might try to construct
these solutions~: by directly writing the appropriate ansatz for
D-branes in the $AdS$ space, or by taking an appropriate near horizon
limit of the supergravity solutions for intersecting D-branes. The
problem with the second approach is that fully localised supergravity
solutions for D-brane intersections are usually not known. The only
exceptions are \cite{smith1}-\cite{papado}. Although the simplified
ansatzes that are often used in the literature preserve the right
amount of supersymmetry, they necessarily require that at best one of
the intersecting D-branes is smeared along the worldvolume directions
of the other D-brane\footnote{The only cases where this is not true is
when there are no relative transverse directions for one of the
D-branes, namely for a D-brane sitting inside an higher dimensional
D-brane or when there are no overall transverse directions.}.  These
are the so-called partially localised solutions.  However, even in these
cases, the solutions take a very complicated form~\cite{pope}. Some
simplification occurs if one replaces the harmonic function for a
smeared brane by its near horizon expression~\cite{youm}. However,
these solutions then describe D-branes in the near horizon geometry of
the smeared D-branes.

Taking the Penrose limit simplifies some of the problems that are
present for $AdS$ spaces. For instance, while the CFT construction of
D-branes in $AdS$ spaces is rather complicated
\cite{bachas,malda3,schomerus} it is much easier to treat this problem
in pp-wave backgrounds \cite{billo, alish}. So one might hope that
constructing D-brane supergravity solutions in these backgrounds is
also more tractable. Performing this exercise would be useful, since
it could teach us how to attack the similar problem in
$AdS$ spaces.

As in $AdS$ spaces, there are two strategies that one can adopt to
find the solution. In one approach, one starts with the solution for a
D-brane in an $AdS$ space and takes its Penrose limit.  This technique
has been used recently in \cite{kumar,singh}, starting from the
supergravity solution~\cite{papado} for $D5/M5$ branes wrapping the
$AdS_3 \times S^3$ space.  The resulting configurations correspond to
localised solutions for $D5/M5$-branes in pp-wave
backgrounds.\footnote{Note that it does not make much sense to take
the Penrose limit of the delocalised solutions of \cite{youm}, since
they do not correspond to (delocalised) D-branes deforming an $AdS$
space.}.  Related work on the D/M-branes in these pp-wave backgrounds
has been done in \cite{clp1}.

The pp-wave which will be considered in this paper is characterised by
a non vanishing 5-form RR flux and arises as the Penrose limit of the
$AdS_5 \times S^5$ space. Since at present there are no known
solutions describing  D-branes in this space, one cannot use the
previous approach. Therefore, the method we will adopt is to directly
write an ansatz for the D-brane in the pp-wave background.

Recently, in \cite{skenderis}, various embeddings of D-branes in the
Hpp-wave geometry were explored using the D-brane probe
approach. There are three different families of
D-branes in Hpp-waves~: {\it longitudinal D-branes} for which the pp-wave
propagates along the worldvolume of the D-brane, {\it transversal D-branes} for which
the pp-wave propagates in a direction transverse to the D-brane but the
timelike direction is along its worldvolume and {\it instantonic
D-branes} for which both the direction in which the pp-wave propagates and
the timelike direction are transverse to the D-brane. In this paper we
consider supergravity solutions for longitudinal D-branes only.

In order to write an ansatz for the D-branes in the Hpp-wave
background, we first study in 
section~\ref{geometry} the geometry of this space and
various brane embeddings.  One important property is that, due to the
\emph{nonisotropy} of the space, D-branes with different orientations
can preserve different amounts of supersymmetry.  Another point is
that the pp-wave is \emph{homogeneous} space, hence all points in the space are
equivalent. However, this property is not manifest in most coordinate
systems, and naively it looks like the coordinate origin is a
physically distinct point. The homogeneity in particular implies that
all types of branes can pass through any point in space at a given
moment in time. It can happen, however, that their embedding when
passing through the origin for example, looks much simpler than when
passing through an arbitrary point in space. We demonstrate this
explicitly on the example of $1/2$ and $1/4$-BPS D-branes.

The first step of writing the ansatz,  consists in identifying 
``natural'' coordinates in which the ansatz looks the simplest.  By
rewriting the embedding of different probe D-branes in various
coordinate systems, we first identify in section~\ref{geometry} the Brinkman
coordinates as the natural coordinates for describing $1/4$-BPS and
nonsupersymmetric D-branes. In section~\ref{Ansatz} we write an ansatz
in these coordinates.  We also argue that the natural coordinates for
the $1/2$-BPS D-branes are the Rosen coordinates.

One of the main characteristics and perhaps limitations of our ansatz
is that the metric is diagonal in Brinkman coordinates.  This
property will force us to delocalise the \emph{supersymmetric}
solutions along some directions transverse to the brane when solving
the equations of motion. However, these restrictions have to be
imposed only on the harmonic function characterising the D-brane, and
not on the function characterising the pp-wave. Hence, all our
solutions asymptotically tend to the \emph{unmodified} Hpp-wave.
Also, despite the simplicity of the ansatz, the \emph{non-supersymmetric}
solutions that we  find are fully localised.

In section~\ref{susies} we analyse the supersymmetry preserved by our
ansatz according to the orientation of the D-brane and we find
agreement with the CFT/probe brane analysis.  In
section~\ref{solutions} we solve the supergravity equations and give
the explicit form of the solutions, including the non-supersymmetric
ones.  Those exhibit a repulsive behavior close to the D-brane, even
for the non-singular D3-brane solution.  The important issue of
stability of these branes is currently under investigation using the
probe brane/CFT approaches.  We conclude with comments on open
questions. Finally, we attach two appendices. In the first one, we present
some technical details of the calculations. In the second appendix, we
repeat some of the D-brane probe analysis of~\cite{skenderis} with 
special emphasis on the nonsupersymmetric branes, in order to make a
clear parallel between the supersymmetry analysis from the probe brane
and supergravity points-of-view.

\section{Setting up an ansatz}

\subsection{The geometry of the Hpp-wave and D-brane embeddings}
\label{geometry}

To set up an ansatz suitable to obtain the supergravity solution describing
a D-brane in the Hpp-wave background of~\cite{blau}, it is important
to first understand the geometry of this space. The metric part of the
solution has isometry group $SO(1,1) \times SO(8)$ which is further
broken down to $SO(1,1)\times SO(4) \times SO(4)$ by the presence of a
null RR 5-form flux. As mentioned in the introduction, this space is
nonisotropic and homogeneous.

The main difficulty in constructing these supergravity solutions
consists in identifying a coordinate 
system where the description of the D-brane is the simplest. This is
similar to the problem that one would face if one would only know
the Minkowski space in spherical coordinates and try to describe
flat D-branes in these coordinates. Cartesian coordinates are the
natural coordinates to describe infinite D-branes in flat space. So
the question that one should first ask is what are the analogues of
the Cartesian coordinates for D-branes in pp-wave backgrounds?  The
answer to this question is more complicated than in flat space, and as
we will now see it depends very much on what kind of D-branes one
considers.

Some possible supersymmetric embeddings of D-branes in the Hpp-wave
space-time have been explored in \cite{skenderis,dabholkar}
using the D-brane probe and the CFT approaches respectively.  Both
analyses were performed in Brinkman coordinates, in which the pp-wave
metric and 5-form flux read
\begin{eqnarray}
  \label{KG_10}
  ds^{2} &=& 2du\left( dv \,+\, Sdu\right)
        \,-\, d\vec{z}_{(8)}^{\,2} \hspace{.2cm},\hspace{.2cm}
     S \ =\ -\textstyle{\frac{W^{2}}{32}} z^{\mu} z_{\mu} \; , \nonumber \\
  F_{(5)} &=& W\ du\wedge\left( 
                  dz^1 \wedge dz^2 \wedge dz^3 \wedge dz^4 + dz^5 \wedge dz^6 \wedge dz^7 \wedge dz^8 
              \right) \; ,
\end{eqnarray}
where the directions transverse to the pp-wave are denoted by $z^{\mu}=
z^1,\ldots ,z^8$.  If we embed a D$p$-brane in this background in such
a way that the pp-wave propagates along its worldsheet,
the worldvolume coordinates split into three sets~: the ``lightcone
coordinates''\footnote{Strictly speaking $(u,v)$ coordinates are not
lightcone; only the direction $v$ is null, while $u$ is timelike for $S
\neq 0$.} $u$ and $v$, $m$ coordinates along the first $SO(4)$
subspace $z^1,\ldots,z^4$ and $n$ coordinates along the second $SO(4)$
space $z^5,\ldots,z^8$. For a D$p$-brane ($m+n=p-1$) with such
orientation we adopt the notation of \cite{skenderis} and denote such
embeddings with $(+,-,m,n)$.

One of the conclusions of \cite{skenderis} is that there are three families
of longitudinal, infinite, flat (in Brinkman coordinates) D-branes with no worldvolume
fluxes that sit at the \emph{origin} of the pp-wave.
The number of supersymmetries that they preserve
depends on their orientation with respect to the isometry group
$SO(4)\times SO(4)$~:
\begin{itemize}
\item $1/2$-BPS D-branes with the embedding $(+,-,m+2,m)$, for  $m=1,\ldots,4$,
\item $1/4$-BPS D-string with the embedding $(+,-,0,0)$,
\item non-supersymmetric D-branes with the embedding $(+,-,m,m)$
for $m=1,2,3$.
\end{itemize}
In \cite{skenderis}, it was also noticed that to \emph{rigidly} move
the first type of D-brane away from the origin and to preserve the
same amount of supersymmetries, one has to turn on a worldvolume flux
whose value is equal to the distance of the D-brane from the
origin. By rigidly we mean here that the D-brane is located at a
constant transverse position~$y^{\hat{a}}$. For any other value of the
flux, including zero, a D-brane that is rigidly sitting away from the
origin in Brinkman coordinates is only $1/4$-BPS.

However, it is important to realise that since the space is
homogeneous, there is nothing physically special about the origin of
the space. This means that all listed $1/2$-BPS and $1/4$-BPS branes
can pass through \emph{any} point in the transverse space at a given
moment in time. At each such point there is an \emph{infinite} number
of $1/2$-BPS and an \emph{infinite} number of $1/4$-BPS D-branes,
labelled by different worldvolume fluxes.  However, the embedding of
many of these D-branes looks quite complicated when they are not
positioned at the coordinate origin. We will see this explicitly on
the example of the half and quarter-BPS D-branes with no worldvolume
fluxes. In the remaining of the paper we only consider D-branes with
no worldvolume fluxes.

From the D-brane probe analysis we have learnt that $1/4$-BPS D-branes
have a simple description in Brinkman coordinates, independent of
their position with respect to the origin.  This is an important
property, since in order to find supergravity solutions for the simple
ansatz that we use, we are forced to smear several of these solutions
in some directions. This is the same type of restriction that one
faces when constructing supergravity solutions for intersecting
D-branes, with a simple diagonal ansatz \cite{youm,gomberoff}.
The smearing procedure physically means that one is constructing an
array of D-branes of the same type with an infinitesimally small
spacing.  However, the probe brane results tell us that, unless we
turn on additional bulk fluxes (sourced by the worldvolume fluxes of
the $1/2$-BPS D-branes), a periodic array of \emph{rigid} D-branes in
\emph{Brinkman} coordinates with orientation $(+,-,n+2,n)$ will be
only one quarter supersymmetric!  In conclusion, Brinkman coordinates
are coordinates in which one expects the (smeared) supergravity
solution for the $1/4$-BPS D-branes to look the simplest.

What about $1/2$-BPS D-branes?  As we will see the $1/2$-BPS D-brane
located away from the origin of the Brinkman coordinates has a
complicated shape (see equation~(\ref{embed})). Hence, to describe an
array of these D-branes in Brinkman coordinates, it is not clear what
ansatz one should use.  In Rosen coordinates on the other hand, where
the homogeneity property of the space transverse to the pp-wave is
manifest, these $1/2$-BPS D-branes look very simple, independent of
the position with respect to the origin. In Rosen coordinates, the
metric and the 5-form of the pp-wave take the form
\begin{eqnarray}
\label{rosen}
ds^2 &=& 2 d \tilde{u} d \tilde{v} - \cos^2 (\alpha\tilde{u})
(d\vec{\tilde{z}})^2 \nonumber \\ 
F_{[5]} &=& W \cos^4 (\alpha\tilde{u})\, d\tilde{u} \wedge
\left(d\tilde{z}^1 \wedge d\tilde{z}^2 \wedge d\tilde{z}^3 \wedge
d\tilde{z}^4 + d\tilde{z}^5 \wedge d\tilde{z}^6 \wedge d\tilde{z}^7
\wedge d\tilde{z}^8 \right) \; ,  
\end{eqnarray}
where we have defined $\alpha=W/4$.  These are related to the Brinkman
coordinates~(\ref{KG_10}) by the change of variables
\begin{equation}
\label{coord-change}
u = \tilde{u}, \qquad v = \tilde{v} - \frac{\alpha}{4} \tilde{z}^2
\sin(2 \alpha \tilde{u}), \qquad z = \tilde{z} \cos(\alpha \tilde{u})
\, . 
\end{equation}
If we denote with $y^{\hat{a}}$ the directions transverse to the D-brane, we
see that the $1/2$-BPS D-brane sitting at a position $y^{\hat{a}}=0$ in the
coordinate system~(\ref{KG_10}) has an embedding $\tilde{y}^{\hat{a}} =
0$ after the change of coordinates. However, we can now shift the
origin of the coordinate system by an arbitrary vector $\vec{z}_{0}$~:
$\tilde{z}^\mu \rightarrow \tilde z^{\mu} + z_{0}^\mu$ for
$\mu=1,\ldots,8$, and this $1/2$-BPS D-brane will sit at a position
$\tilde{y}^{\hat{a}}= y_{0}^{\hat{a}}$. Hence, we see that in Rosen
coordinates, $1/2$-BPS D-branes always have a ``flat'' embedding,
independently of the choice of an  origin.

To determine the shape of this D-brane in the
coordinates~(\ref{KG_10}), we have to perform a further change of
coordinates~(\ref{coord-change}) with angle~$\tilde{u}$. This gives us
an embedding of the $1/2$-BPS D-branes in coordinates~(\ref{KG_10})
away from the origin of the space; the $p+1$-dimensional surface that
they describe is defined by the following $9-p$ constraints
\begin{eqnarray}
\label{embed}
y^{\hat{a}} = {y}^{\hat{a}}_{0}\cos(\alpha \tilde{u})\, .
\end{eqnarray}  
So, although $1/2$-BPS D-branes that pass through the
origin in Brinkman coordinates look ``flat'', once we
move them away from the origin, their shape is much more
complicated\footnote{The
procedure that we have just 
described is completely analogous to what one has to do in Minkowski
space to change the origin of a cylindrical system of
coordinates. One first passes to Cartesian coordinates, shifts the
origin of the space and then goes back to cylindrical coordinates.}.
The shapes of $1/2$  and $1/4$-BPS D-branes in Brinkman and Rosen coordinates
are depicted in figure~\ref{shapes}. 
\FIGURE[th]{
\vbox{%
\hbox{
  \vbox{\hbox{\quad\quad\quad\includegraphics[angle=-90,width=.3\textwidth]{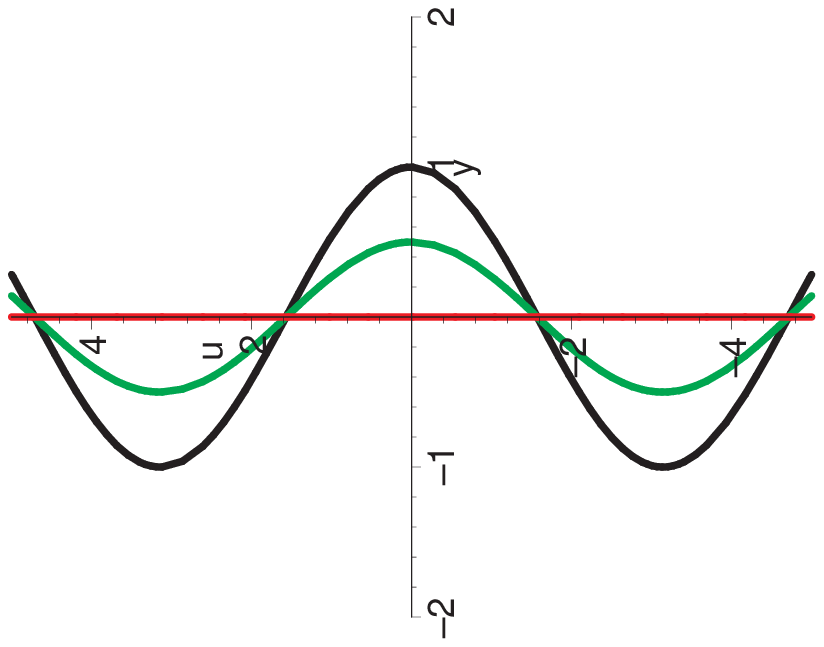}}\vskip1ex
        \hbox{\small $1/2$-BPS branes in Brinkman coordinates}}\quad\quad
  \vbox{\hbox{\quad\quad\quad\includegraphics[angle=-90,width=.3\textwidth]{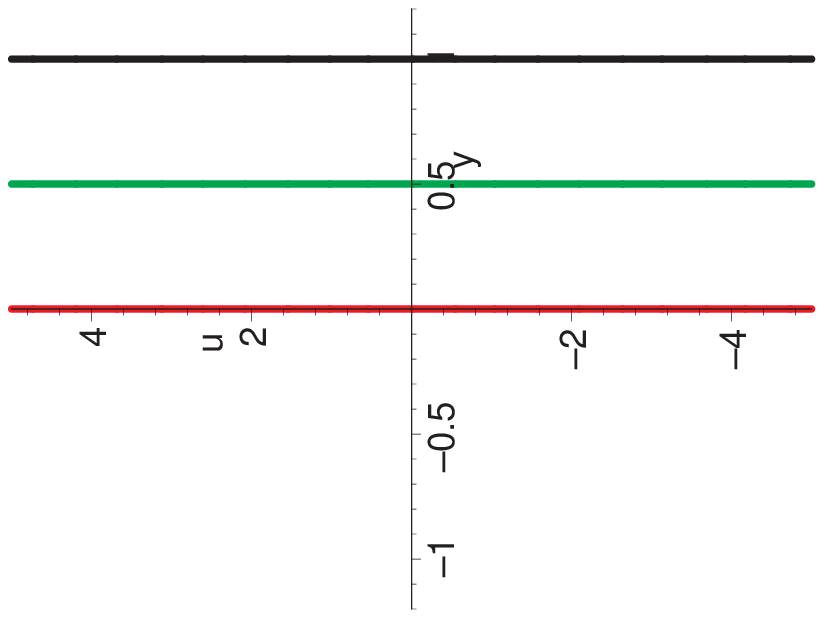}}\vskip1ex
        \hbox{\small $1/2$-BPS branes in Rosen coordinates}}}
\vskip 3ex
\hbox{
  \vbox{\hbox{\quad\quad\quad\includegraphics[angle=-90,width=.3\textwidth]{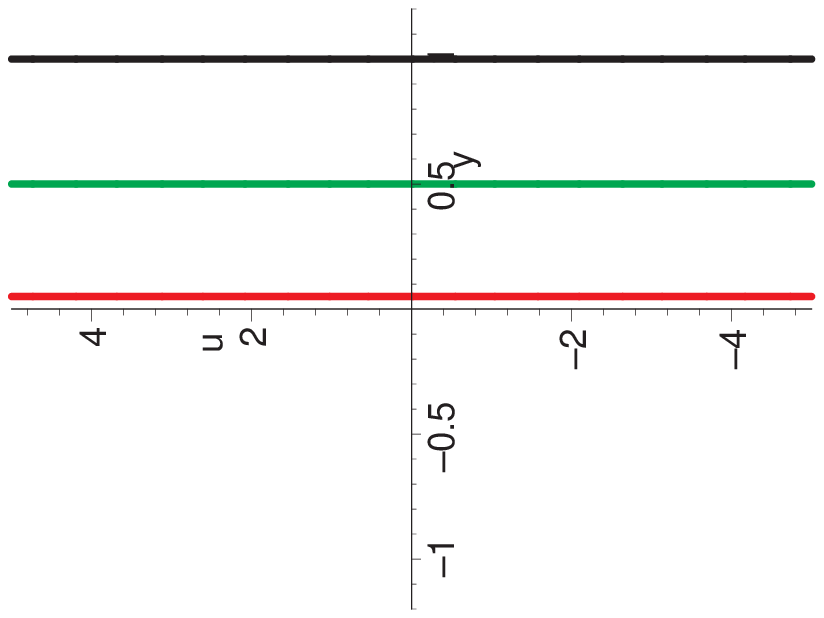}}\vskip1ex
        \hbox{\small $1/4$-BPS branes in Brinkman coordinates}}\quad\quad
  \vbox{\hbox{\quad\quad\quad\includegraphics[angle=-90,width=.3\textwidth]{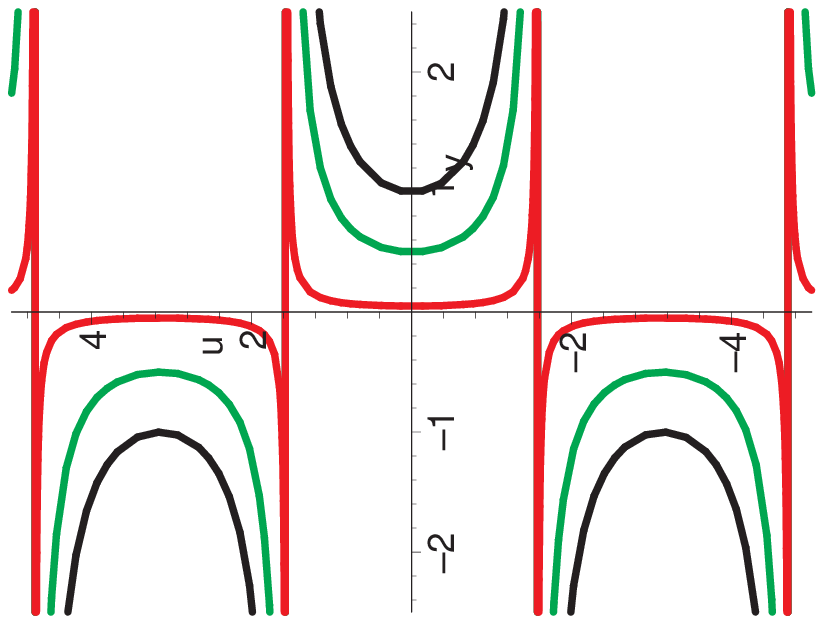}}\vskip1ex
        \hbox{\small $1/4$-BPS branes in Rosen coordinates}}}}
\caption{The shapes of $1/2$ and $1/4$-BPS D-branes in Brinkman and
Rosen coordinates. The vertical axes correspond to the time-like coordinate
$u$, while the horizontal axes correspond to the directions transverse
to the brane.}
\label{shapes}
}

This preliminary discussion leads us to expect that the natural
coordinates one should use to write the (smeared) supergravity ansatz
for the $1/2$-BPS D-branes are Rosen coordinates. However, a quick
inspection reveals that the ``naive'', general \emph{diagonal}
ansatz that one can write in these coordinates does not seem to
allow for the $1/2$-BPS solutions. We are currently investigating 
more general ansatzes in these coordinates. 

Finally, it is interesting to note that the worldsheets of $1/2$-BPS
D-branes are geodesic submanifolds (the extrinsic curvature of the
worldsheet vanishes).  This is similar to situation in Minkowski
space where maximally supersymmetric D-branes are flat sheets spanned by
straight lines, i.e.~Minkowski geodesics. The main difference is
that geodesics in this space are not straight lines in \emph{Brinkman}
coordinates. They are curves, depicted in figure~\ref{shapes}a and
given by equations
\begin{equation} 
z = A \cos(\alpha u) \, , \qquad v = B u + C \sin(2\alpha u ) \, ,
\end{equation}
where $A$, $B$ and $C$ are constants determined in terms of lightcone
momenta and energy. Comparing this expression
with~(\ref{coord-change}) we see that going from Brinkman to Rosen
coordinates means that we are passing to a coordinate system with
the coordinate grid given by the geodesics of a point particle (up to
a rotation in the $u$-$v$ plane). It is now clear why a $1/2$-BPS
D-brane in Rosen coordinates has a flat embedding, independent of the
point through which it passes.

Note also that the fact that \emph{all} geodesics starting at the
origin intersect at $u= {\pi \over 2\alpha}$ implies that there are 
apparent coordinate singularities at $u=\frac{\pi}{2\alpha} + {n \pi
\over \alpha}$, $(n \in \mathbb{Z})$ in Rosen coordinates.

Finally, before writing the ansatz, to set up our conventions, we give
the action and the supersymmetry variations that we will need. The
relevant part of the type IIB action is
\begin{equation}
\label{action}
S = \int dx^{10} \sqrt{g} \bigg{\{} e^{-2 \phi} \bigg{(} R -4 (\partial \phi)^2 \bigg{)}  
+ {1\over 2} (\partial C_{[0]})^2 + {1\over 4 \cdot 3!}(F_{[3]})^2 +  {1\over 4 \cdot 5!} (F_{[5]})^2 + {1\over 4 \cdot 7!} (F_{[7]})^2 \bigg{\}} \, .
\end{equation}
where the  5-form field strength is self-dual and satisfies the
Bianchi identity $dF_{(5)}=0$. The corresponding supersymmetry variations are 
\begin{eqnarray}
\label{susy}
\delta \Psi_{\mu} &=& D_{\mu} \epsilon +  {1 \over 16} e^{\phi}
\bigg{(} 2 \slashed{\partial} C_{[0]} (i\sigma^2) + {1\over 3!}
\slashed{F}_{[3]} (\sigma^1) + {1\over 5!} \slashed{F}_{[5]} (i
\sigma^2) + {1\over 7!}\slashed{F}_{[7]} (\sigma^1) \bigg{)}
\gamma_{\mu} \epsilon  \, \nonumber \\ 
\delta {\chi} &=& \slashed{\partial} \phi\, \epsilon + {1\over 4}
e^{\phi}\bigg{(} -4 \slashed{\partial} C_{[0]} (i \sigma^2) -{1\over
3!} \slashed{F}_{[3]}(\sigma^1) + {1\over 7!} \slashed{F}_{[7]}
(\sigma^1) \bigg{)} \epsilon \, , 
\end{eqnarray}
where we have used the doubled formulation of supergravity with both electric
and magnetic RR fields, except for the 0-form~$C_{[0]}$. Therefore, for the 
$D7$-branes, we will express our results using~$C_{[0]}$.

\subsection{The ansatz for quarter-BPS D-branes}
\label{Ansatz}
Motivated by the analysis in the previous section, we use Brinkman
coordinates to write an ansatz for $1/4$-BPS and nonsupersymmetric
D-branes.  For the metric part of the ansatz, we write a simple standard
metric for a superposition of D-branes with pp-waves.
For a pp-wave propagating in the directions $u$ and $v$ along the
worldvolume of the D-brane this ansatz reads
\begin{equation}
\label{ansatza}
ds^2 = H(y,y')^{-{1\over 2}} \bigg{(} 2 du(dv + S(x,x',y,y')du) - d\vec{x}^{\,2}  - d\vec{x}'^{\,2} 
\bigg{)} - H(y,y')^{1\over 2} (d\vec{y}^{\,2} + d\vec{y}'^{\,2})\, . 
\end{equation}
The metric is given in the string frame, and the D-brane worldvolume
coordinates are ($u,v, x^i = x^1,\ldots, x^m, x'^{I}=
x'^1,\ldots,x'^n$), while directions transverse to the D-brane are
$(y^a = y^1,\ldots,y^{(4-m)}, y'^A = y'^1,\ldots,y'^{(4-n)})$. Note
that the function $H$ characterising the D-brane is at this stage
allowed to depend on all transverse coordinates~\mbox{$y$, $y'$}.

The components of the Ricci tensor for this metric are given in
appendix~\ref{app1}. An important observation is that all components
of the Ricci tensor, save $R_{uu}$, are independent of the
function~$S$ appearing in the metric. The easiest way to ensure
that the equations of motion are satisfied is then to keep the RR field and
the dilaton coming from the D-brane unchanged. The components of the
5-form flux of the pp-wave which are kept nonzero are the same as in the
absence of the D-brane.  Hence, we make the following ansatz for the
RR field strength and the dilaton
\begin{eqnarray}
\label{ansatzb}
F_{[p+1]} &=& du \wedge dv \wedge dx^1 \wedge \cdots \wedge dx^m \wedge dx'^1 
 \cdots \wedge dx'^n \wedge dH^{-1} \, , \\
\label{ansatzc}
F^{\rm w}_{[5]} &=& F_{[5]}^{(1)}  + * F_{[5]}^{(1)} \, , \nonumber \\
                  F_{[5]}^{(1)} &=& W(z^{\mu}) \ du\wedge\left(  
                  dx^1 \wedge \cdots \wedge dx^m \wedge dy^1 \wedge
                  \cdots \wedge dy^{(4-m)} \right) \, , \\
\label{ansatzd}
e^{\phi} &=& H^{3-p \over 4}   \, ,
\end{eqnarray}
where $'*'$ in $F_{[5]}$ denotes the Hodge duality with respect to the
metric~(\ref{ansatza}) and $W(z)$ is an undetermined function which
can depend on all directions transverse to the pp-wave.  Also, in the
case of the $D3$ brane, one has to add to the form~(\ref{ansatzb}) its
Hodge dual.  Next, we use the fact that $F_{[5]}$ has to satisfy the
Bianchi identity. This condition, supplemented with the self-duality
of $F_{[5]}$, is enough to ensure that the equation of motion for
$F_{[5]}$ is satisfied. The requirement that $F_{[5]}$ is closed implies
the following set of equations
\begin{eqnarray}
\label{first}
\partial_{x'} W &=& 0 \, , \quad \quad \partial_{y'} W = 0 \, , \\  
\label{second}
\partial_{x} (W H^{{(m-n) \over 2}}) &=& 0 \, , \quad \quad \partial_{y} (W H^{{(m-n)\over 2}}) = 0 \, .
\end{eqnarray}
The first two conditions lead to the restriction $W=W(x,y)$, while the
third one, when supplemented with $H=H(y,y')$, shows that
$W=W(y)$. Finally, the last condition implies 
\begin{eqnarray}
\label{H}
H &=& (W^{-1}(y)\hat{H}(y'))^{2\over (m-n)}  \, , \quad  \quad \quad \quad \quad \, \, \, (m\neq n) \\
\label{H1}
H &=& H(y,y') \, , \quad  W = const. \, , \quad \quad \quad \quad \, \, \,  (m=n) ,
\end{eqnarray}
where $\hat{H}$ is an arbitrary function of $y'$. When $m \neq n$, the
D-brane ``harmonic'' function has a multiplicative dependence on the
transverse coordinates. Usually, this kind of functional dependence
cannot be matched into a delta function brane source~\cite{tseytlin},
and hence one is then forced to drop dependence of $H$ on $y$ or $y'$
and consider only a delocalised solution. We will show now that
consistency of our ansatz with the equations of motion also implies
such smearing.

Let us now consider the constraints coming from the remaining
supergravity equations. Inspection of the dilaton equation of
motion~(\ref{eom-dil}) shows that the dilaton does not couple to the
pp-wave function $S$ and that it is not modified by its 5-form
flux. This equation implies that $H$ is a harmonic function in the
space transverse to the D-brane.

The equations of motion for the RR fields and the Bianchi identities
can be all written in a nice compact form given in the appendix,
see~(\ref{BI-RR}). Since we are looking for supergravity solutions
with vanishing NSNS B-field, we have to ensure that the
Chern-Simons (CS) term in the Bianchi identity for $H_{[7]}$ vanishes
\begin{eqnarray}
\label{H7}
dH_{[7]} = - \frac{1}{2} \, {}^{\star} F \wedge F   \, .
\end{eqnarray}
It easy to check that, with our ansatz, the right hand side is zero
for the $D1$, $D3$ and $D7$-branes. For the $D5$-brane, 
the following embeddings are possible with the corresponding CS terms
(coming from $*F_{[7]} \wedge F_{[5]}$ term), 
\begin{eqnarray}
(+,-,4,0): && CS \sim  \, \epsilon_{ABCD}\partial_{D}(H^{-1}) dx^A
\wedge dx^B \wedge dx^C \wedge du \wedge dx^1 \wedge dx^2 \wedge dx^3
\wedge dx^4 \,  \nonumber \\
(+,-,3,1): && CS \sim \, (\partial_{y} \,  H) dy'^1 \wedge dy'^2
\wedge dy'^3 \wedge du \wedge dx^1 \wedge dx^2 \wedge dx^3 \wedge dy
\, \nonumber \\
(+,-,2,2): && CS = \, 0 \, . 
\end{eqnarray}
In the first case, we see that unless the function $H$ is trivial, our
ansatz without NSNS fields is incompatible with the equations of
motion and Bianchi identities. Using the probe brane approach, one can
also check that a D5-brane with a $(+,-,4,0)$ embedding is consistent
only if one allows for non-trivial worldvolume fluxes without or
with non-constant transverse scalars fields. This last possibility
arises as the Penrose limit of a D5-brane corresponding to the baryon
vertex in the gauge theory dual~\cite{witten}. In this case, one
expects to have spikes due to fundamental strings stretched between
the D5-branes wrapped on $S^5$ and the D3-branes whose near horizon
geometry generates $AdS_5 \times S^5$. The supergravity manifestation
of this phenomenon is the presence of non-trivial NSNS fluxes induced
by the contribution of the RR fields to the CS couplings which
presumably survives after one takes the Penrose limit. However, we
will not discuss this kind of solutions.

The second embedding is consistent with our ansatz iff the function
$H$ is independent of the direction $y$ in the first $SO(4)$ subset
and transverse to the D5-brane. Hence, we see that the restricted form
of ansatz forces us to \emph{smear} the D-brane configuration along
this direction.  Finally, the last embedding does not impose any
restriction on $H$.  In the absence of CS terms, the RR Bianchi
identities imply that the (smeared) function $H$ is harmonic in the
space transverse to D-brane. This is compatible with the dilaton
equation.

Finally, let us analyze the Einstein equation.  As already pointed
out, the pp-wave function $S$ appears only in the ${uu}$ component of
the Ricci tensor; all other components are the same as for a D-brane
in a flat background. Note that the RR flux of the pp-wave contributes
only to the $T_{uu}^{RR}$ component of the energy momentum
tensor. This can easily be seen by realising that the only component
of the inverse metric which involves the index $u$ is $g^{uv}$; hence,
since $F_{[5]}$ does not contain the index $v$, there is no way in which
one can contract the index $u$.  Therefore, to be consistent with our
ansatz, one must check that there are no extra source terms in other
components of the energy momentum tensor coming from the cross-terms
between the RR 5-form of the wave and the RR fields of the
D-brane. This is obviously true for any D-brane except the
D3-brane. In this case, the possible embeddings are $(+,-,2,0)$ and
$(+,-,1,1)$. In the second configuration, all cross-terms vanish
identically. However, in the first case, there are non-vanishing
cross-terms
\begin{equation} 
T_{uy^1} \sim \partial_{y^2} H \, , \quad \quad \quad T_{uy^2} \sim
\partial_{y^1} H \, . 
\end{equation}
Hence we see that for an $(+,-,2,0)$ embedding, in order to satisfy
the equations of motion,  we are again forced  to smear the function
$H$ in the directions $y^1$ and $y^2$.   

Provided that one imposes this constraint, the only nontrivial
component of the Einstein equation is the $uu$ component
\begin{equation}
\label{eqns_s}
- {1\over 2} S H^{-2}(\partial_{\vec{y}}^2 + \partial_{\vec{y}'}^2) H  + H^{-1} (\partial_{\vec{y}}^2 +  \partial_{\vec{y}'}^2) S + (\partial_{\vec{x}}^2 +  \
\partial_{\vec{x}'}^2) S = {1 \over 2} W^{2} H^{{m-n-2\over 2}} 
\end{equation}
while all the other equations of motion and Bianchi identities reduce
to 
\begin{equation}
\label{eqns_h}
(\partial_{\vec{y}}^2 + \partial_{\vec{y}'}^2) H = 0  \,.
\end{equation}

\section {Supersymmetry analysis}
\label{susies}
We now want to show that our ansatz preserves the same amount of
supersymmetry as predicted by the probe analysis~\cite{skenderis}. To
make the parallel between the supersymmetry analysis in the supergravity and
the probe brane approaches clear, we give some of the probe brane result
in appendix~\ref{app2}.  The smeared solutions in Brinkman
coordinates describe an array of flat $(+,-,m+2,m)$ branes. Therefore,
from the D-brane probe point-of-view \cite{skenderis}, we expect them
to preserve only one quarter of the supersymmetries. The $(+,-,m,m)$
branes should lead to non-supersymmetric solutions, except for the
$(+,-,0,0)$ D-string which should preserve one quarter of the
supersymmetries.

Specialising the supersymmetry variations (\ref{susy}) to our ansatz
gives the following set of equations. First, the dilatino variation
reads
\begin{equation}
\label{dilatino}
\delta \chi = \left(\partial_{a} H\gamma^{\underline{a}} +
\partial_{A} H \gamma^{\underline{A}} \right) 
\left(1 - \gamma^{\underline{uvx^1 \cdots x^m x'^1 \cdots x'^n}} {\cal P}_k\right) \epsilon = 0 \, ,
\end{equation}
where $p=2k+1$. For $k$ even, ${\cal P}_k = \sigma^1$ and for $k$ odd,
${\cal P}_k = i\sigma^2$. Underlined indices correspond to tangent
space indices.  So we see that one has to impose 
the standard projection condition for D-branes in flat space
\begin{equation}
\label{s1}
\gamma^{\underline{uvxx'}} {\cal P}_k \epsilon = \epsilon \, .
\end{equation}
Here we have introduced the short notation
\begin{equation}
\gamma^{\underline{uvxx'}} \equiv \gamma^{\underline{uvx^1 \cdots x^m x'^1 \cdots x'^n}} \, , 
\end{equation}
and similarly for the other combinations of $x$, $x'$, $y$ and $y'$.

Using the identity $(\gamma^{\underline{u}})^2 = 0$, one shows that
the gravitino variation in the  direction $v$ simply reduces to the
condition 
\begin{equation}
\partial_{v} \epsilon = 0 \,  .
\end{equation}
The gravitino variations in the directions $\hat{a}=(a,A)$ read
\begin{equation}
\delta \Psi_{\hat{a}} = \partial_{\hat{a}} \epsilon + {1\over 8} H^{-1} (\partial_{\hat{a}} H) \epsilon  + {1\over 8} W H^{{m-n \over 4}} \gamma^{\underline{uxy}}\gamma_{\underline{\hat{a}}}(i\sigma_2) \epsilon = 0 \, .
\end{equation}
Instead of trying to solve this equation directly, it is instructive
to first consider the integrability condition coming from the
gravitino variation in these directions 
\begin{equation}
\label{intg}
[\delta \Psi_{\hat{a}}, \delta \Psi_{\hat{b}}] = \partial_{[\hat{a}}(W H^{(m-n)\over 4}) \gamma_{\underline{\hat{b}}]} \gamma^{\underline{u}}(i\sigma_{2}) \epsilon = 0 \, .
\end{equation}
We first try to satisfy this condition by imposing that 
\begin{equation}
\label{const.}
W H^{(m-n)\over 4} = const. 
\end{equation}
We will now discuss separately the cases $m\neq n$ and $m=n$.

\vskip10pt

\noindent\underline{$m\neq n$ :}

\vskip8pt 

In this case, using (\ref{H}), the condition (\ref{const.}) implies that
$W(y)^{1/2}\hat{H}(y')^{1/2}$ is a constant. This is true only if $W$
and $\hat{H}$ are constant and therefore, if $H$ is trivial.  In order
to satisfy (\ref{intg}) with a nontrivial $H$, we are forced to impose 
\begin{equation}
\label{cond-u}
\gamma^{\underline{u}} \epsilon = 0 \, .
\end{equation} 
We still have to check whether satisfying the integrability condition
is enough to ensure that the supersymmetry variation of
$\Psi_{\hat{a}}$, 
\begin{equation}
\delta \Psi_{\hat{a}} = \partial_{\hat{a}} \epsilon + {1\over 8} H^{-1} (\partial_{\hat{a}} H) \epsilon = 0 \, ,
\end{equation}
is satisfied.
The solution to this equation is
\begin{equation}
\epsilon = H^{-{1\over 8}} \hat{\epsilon}(u,x,x') \, .
\end{equation}
Using the condition (\ref{cond-u}), the gravitino variation in the
$\hat{\imath} = (i, I)$ directions becomes trivial
$\partial_{\hat{\imath}}
\epsilon = 0 $, while the variation in the $u$ direction reduces to 
\begin{equation}
\label{u-susy1}
\delta \Psi_{{u}} = \partial_{u} \epsilon(u) + {1\over 8} W H^{(m-n-2)\over 4} \gamma^{\underline{uxy}}\gamma^{\underline{v}} {\cal P}_{k} \hat{\epsilon}(u) = 0 \, .  
\end{equation}
Since $\hat{\epsilon}$ does not depend in $y,y'$, this equation
 has a nontrivial solution only if
\begin{equation}
\label{conditions}
W H^{(m-n-2)\over 4} =  W(y)^{m-n+2 \over 2 (m-n)} \hat{H}(y')^{m-n-2 \over 2 (m-n)} = const. \quad ,
\end{equation}
which further implies
\begin{eqnarray}
\label{conditions1}
m-n &=& 2 \, \, \quad \quad W(y) = const. \, \quad \hbox{or} \, \nonumber \\
n-m &=& 2 \, \, \quad \quad \hat{H}(y) = const. 
\end{eqnarray}
In other words, we see that the $(+,-,m,m\pm2)$ embeddings will be
supersymmetric only if we smear the function $H$ in part of the
transverse space. These 1/4-BPS configurations are summarised in
table~(\ref{sum1}).

\vskip10pt

\noindent\underline{$m = n$ :}
\vskip8pt

In this case, the condition (\ref{const.})  implies that $W$ is
constant, which is consistent with  the condition~(\ref{H1}).  
The gravitino variation $\delta \Psi_{\hat{a}}$, reduces to 
\begin{equation}
\delta \Psi_{\hat{a}} = \partial_{\hat{a}} \epsilon + {1\over 8} H^{-1} (\partial_{\hat{a}} H) \epsilon  + {1\over 8} W \gamma^{\underline{uxy}}\gamma_{\underline{\hat{a}}}(i\sigma_2) \epsilon = 0 \, 
\end{equation}
which can now be integrated, yielding the spinor
\begin{equation}
\label{spinor2}
\epsilon = H^{- {1\over 8}} (1 - {1\over 8}W \gamma^{\underline{uxy}} y^a \gamma_{\underline{a}} (i \sigma_2)) \hat{\epsilon}(u,x,x') \, .
\end{equation}
Using this expression in the gravitino variations in
the directions $\hat{\imath}= (i,I)$ give
\begin{equation}
\label{susy-i}
\partial_{\hat{\imath}}\hat{\epsilon}(u,x,x')  + {1\over 8} W H^{-\frac{1}{2}} \gamma^{\underline{uxy}}\gamma_{\underline{\hat{\imath}}} (i\sigma_2) \hat{\epsilon}(u,x,x') = 0 \, . 
\end{equation}
Since $\hat{\epsilon}$ does not depend on $y,y'$ the only way this
equation can be satisfied is if we again impose the
condition~(\ref{cond-u}). Here we recover the same constraint as in
the D-brane probe analysis described in the appendix, where
equation~(\ref{proj2}) is also consequence of the existence of
worldvolume coordinates~$x^{\hat{\imath}}$.

Finally, the supersymmetry variation in the direction $u$ reduces to
\begin{equation}
\label{u-susy}
\partial_{u} \epsilon(u) + {1\over 4} W H^{-{1\over 2}} \gamma^{\underline{xy}}(i\sigma_2) \epsilon(u) = 0 \, ,  
\end{equation}
which obviously does not have any solutions. Hence the $(+,-,m,m)$
embeddings of the D-branes break all supersymmetries! This is precisely
the result predicted by the D-brane probe analysis and  we
see that supersymmetry fails for the same reason in both
approaches. Namely, in both cases it is not possible to simultaneously
satisfy the constraints coming from the supersymmetry/kappa symmetry
conditions in the $u$ and $x^{\hat{\imath}}$ directions. 
 
The previous discussion does not hold if there are no worldvolume
directions $x^{\hat{\imath}}$, namely for the D-string. In this case,
equation~(\ref{susy-i}) is missing and therefore one does not need to
impose the projection~(\ref{cond-u}) anymore.  The variation in the
$u$~direction now becomes
\begin{equation}
\partial_u \hat{\epsilon}(u) - {1\over 2}H^{-{1\over 2}}[ (\partial_{\hat{a}} S) - {1\over 16} W^2 y^a] \gamma^{\underline{u \hat{a}}} \hat{\epsilon}  + {1\over 8}W H^{-{1\over 2}} \gamma^{\underline{y}} [(\gamma^{\underline{uv}} + \eta^{\underline{uv}}) i\sigma_2] \hat{\epsilon} = 0 \, .
\end{equation}
The content of the first bracket gives the BPS equation for $S$,
\begin{equation}
\partial_{\hat{a}} S - {1\over 16} W^2 y^{\hat{a}} = 0 \, ,
\end{equation}
while the content of second bracket, when combined with~(\ref{s1}) implies
the condition  
\begin{equation} 
(1 + \sigma_1) \hat{\epsilon} = 0 \, .
\end{equation}
In conclusion, the D-string preserves exactly one quarter of the
supersymmetries.  Again, let us emphasise that the D-brane probe
discussion is completely parallel.  All the results of this
supersymmetry analysis are summarised in table~\ref{sum1}.

\renewcommand{\arraystretch}{1.5}
\begin{table}[th]
\begin{center}
\begin{tabular}{|c|c|c|}
\hline
brane & embedding & susy \\
\hline
D1 &$(+,-,0,0)$&$ 1/4 $ \\
\hline
D3 & $(+,-,2,0)$ &$1/4$ \\ 
& $(+,-,1,1)$  & $-$ \\
\hline
D5 & $(+,-,3,1)$ & $1/4$ \\
& $(+,-,2,2)$ & $-$ \\
\hline
D7 & $(+,-,4,2)$ &$1/4$ \\
& $(+,-,3,3)$  & $-$ \\
\hline
\end{tabular}
\end{center}
\caption{Amount of supersymmetry preserved by the various D-brane
supergravity solutions.}
\label{sum1}
\end{table}

\section{Solving the field equations}
\label{solutions}
In this section we solve the equations of motion~(\ref{eqns_s})
and~(\ref{eqns_h}). For all solutions (except the D-string), the
presence of the D-brane modifies the function $S$ which characterises
the pp-wave. On the other hand, for our ansatz, the field equations
imply that the function $H$ (which specifies the D-brane) is
completely unmodified by the presence of the pp-wave. Therefore, this
ansatz does not catch the \emph{back-reaction of the pp-wave on the
D-brane}. For a generic embedding of the D-brane, one expects that
the (fully localised) D-brane is modified by the pp-wave. However, as our
fully localised, nonsupersymmetric solutions demonstrate, this does not
have to hold for some specific embeddings.

We will divide the discussion of the solutions according to the
number of supersymmetries that they preserve,
describing first the $1/4$-BPS D-branes; we can split them into two
subclasses~: unsmeared solutions ($D1$ and $D7$) and smeared ones ($D3$ 
and $D5$). Then, we turn to the integration of equations~(\ref{eqns_h})
and (\ref{eqns_s}) for the \emph{non-supersymmetric}
$(+,-,m,m)$-embeddings.

\subsection{Supersymmetric solutions}

{\it D-string~:} \vskip10pt The D-string has Neumann boundary
conditions along the lightcone coordinates $u$ and $v$ and Dirichlet
boundary conditions along the space transverse to the pp-wave. In this
case, equations~(\ref{eqns_s}) and~(\ref{eqns_h}) are fully decoupled
and hence the pp-wave and the D-string are completely insensitive to
the presence of each other.  Explicitly, the solution is given by
\begin{eqnarray}
H &=& 1 + {Q\over (y^2 + y'^2)^3} \nonumber \\
S &=& {1\over 32} W^2 (y^2 + y'^2) \, , \quad \quad W = const. \, .
\end{eqnarray}

\noindent{\it Smeared D-branes:}
\vskip10pt
The supersymmetry analysis performed above leads to the conclusion that 
$D3$ and $D5$-branes with the respective embeddings $(+,-,2,0)$ and
$(+,-,3,1)$ are supersymmetric only if the function $H$ does not
depend on the coordinates $y$ transverse to the brane that belong to
the first $SO(4)$ subspace. The harmonic functions $H$ in this case
are  
\begin{equation}
D3~: \quad H = 1 + {Q\over |y'|^2} \, , \quad \quad\quad \quad D5~:
\quad H = 1 + {Q \over |y'|} \, . 
\end{equation}
We are interested in solutions which asymptotically, far away from the
D-brane, reproduce the Hpp-wave vacuum~(\ref{KG_10}). To get a
solution with this property, one can choose to decompose $S$ as the
sum of a function which depends on $x$ and a function which depends on
$y$~
\begin{equation}
S(x, x',y, y')=S_x(|x|) + S_{x'}(|x'|) + S_y(|y|)+ S_{y'}(|y'|)  \, .
\end{equation}
With this ansatz the $x$-dependent and $x'$-dependent parts of the
equation (\ref{eqns_s}) must vanish separately. Imposing that the
solution reduces asymptotically to (\ref{KG_10}) fixes the functions
$S_x$ and $S_{x'}$,
\begin{equation}
S_x + S_{x'} = \frac{1}{32} W^2 (|x|^2 + |x'|^2) \, ,
\label{sx}
\end{equation}
up to a homogeneous solution of the Laplace equation.
This part would correspond to a pure gravitational pp-wave propagating
along the worldvolume of the D-brane (and smeared in the transverse
directions). Since we are not interested in  pure gravitational
pp-waves, we have chosen to set this part to zero in (\ref{sx}). For
the same reasons we will, in what follows, always ignore this kind of
freedom when solving the equation for $S$.

To write an ansatz for the $y, y'$-dependent parts of the metric, we
will assume that the pp-wave is modified only in the directions
transverse to the brane and the directions of smearing.\footnote{Since
the direction along which we smear is not a direction of isometry of
the wave (and since gravity is a highly nonlinear theory), it is not
a priori clear that this assumption should hold. However, we will see
that it leads to a physically acceptable solution.}  Together with the
asymptotic conditions, this assumption leads to the following formula
\begin{equation}
S_{y} + S_{y'} =  {1\over 32} W^2 (|y|^2 + |y'|^2) + f(|y'|^2) \, .
\label{sy}
\end{equation}
where $f$ is determined by plugging (\ref{sx}) and (\ref{sy}) into
the field equation (\ref{eqns_s}). The solutions are given by
\begin{eqnarray}
D3~: \qquad f &=&   {3\over 16} Q W^2 \ln|y'|  \, , \nonumber \\
D5~: \qquad f &=&   {1\over 8 } Q W^2 |y'| \, .
\end{eqnarray}
Let us now examine some of their properties.
First, since $\ln |y'|$, $|y'|$ grow slower than the quadratic terms
in the pp-wave $S$ function, it is clear that both solutions
asymptotically reduce to~(\ref{KG_10}), as required. So although it
looks like the deformation of the pp-wave due to the D-brane grows
(and blows up) at infinity, this is just an artifact of the
coordinates in which the pp-wave is written. 

Second, it seems that in the case of the $D3$-brane, the corrections
to the pp-wave $S$ function blow up as we approach the D-brane. To
explore whether this is indeed true, we look for possible invariant
quantities that one can construct from the metric.  The Ricci scalar
is finite everywhere and independent of $S$. Other quantities (like
$R_{\mu\nu\xi\eta}R^{\mu\nu\xi\eta}$, $R_{\mu\nu}R^{\mu\nu}$) are also
independent of $S$, and diverge for small $r$. The only quantity which
depends on $S$ is given by $R_{\mu\nu}\xi^{\mu}\xi^{\nu}$, where
$\xi=\partial_u$ is the timelike Killing vector field. By looking at
the $R_{uu}$ component in the appendix, one can easily see that this
scalar also diverges for small $r$. Note that the divergent term in
this scalar originates from corrections to the $S$ functions.
 
We now want to see whether these quantities diverge along 
timelike/null geodesics in finite proper time.
So let us consider the geodesics for a radially infalling particle. To
write down the equation of the geodesics, we use of the fact that
$\xi=\partial_u$ and $\eta=\partial_v$ are Killing vectors with 
corresponding conserved quantities~: the light-cone energy $p_{+}$ and the
light-cone momentum $p_{-}$, 
\begin{equation}
(\xi)_{\mu} p^{\mu} = H^{-{1\over 2}}(\dot{v} + 2 S \dot{u}) = p_{+}
\, , \quad \quad (\eta)_{\mu} p^{\mu} = H^{-{1\over2}} \dot{u} = p_{-}
\, . 
\end{equation}
Here dots denote  derivatives of the coordinates with respect to the
proper time.\footnote{We are choosing the parameter along the geodesic
to be equal to the proper time.}
Using these quantities, one can eliminate $\dot{u}$ and $\dot{v}$ from
the action of the point particle. 
Furthermore, one can check that it is consistent with the equations of
motion to set $x=x'=y=0$. Plugging this back into the metric we get
\begin{equation} 
\epsilon = p_{+}p_{-}H^{1\over2} - 2 p_{+}^2 H^{1\over2} S - H^{1\over2}
\dot{y}^2 \, ,
\end{equation}
where $\epsilon=-1,0,1$ for spacelike, null and timelike geodesics respectively.
Hence the equations of motion for the radially infalling particle
reduce to 
\begin{equation}
\label{pt1}
(\dot{y}')^2 = {-\epsilon \over \sqrt{1 + {Q \over (y')^2}}} + 2 p_{-} (p_{+} -
{1\over 32}W^2 p_{-} y'^2 - {3\over 16}QW^2 p_{-}\ln|y'|)\, .
\end{equation}
We see that by setting $Q=0$ one recovers the equation for the linear
harmonic oscillator, as expected. If we have $W=0$, the previous
equation can be integrated for small $y'$, leading to conclusion that
a radially infalling particle can reach the singularity in finite
proper time.
A direct inspection of~(\ref{pt1}) for small $y'$ reveals that turning
on $W$ makes the proper time in which the particle reaches the singularity
shorter. In conclusion, the presence of the pp-wave strengthens the
attraction of the singular smeared D3-brane solution.

\vskip10pt
\noindent{\it $D7$-brane:}
\vskip10pt

Finally, we can also write down a supergravity solution which
describes a D7-brane with the $(+,-,4,2)$ embedding.  In this case,
the smearing of the harmonic
function $H$ is not required. Hence, one gets a fully localised solution
\begin{eqnarray}
H &=& 1 + Q \ln | y' | \, \nonumber \\
S &=& {1\over 32} W^2 (x^2 + x'^2 + y^2 + y'^2 ) - {2\over 32} W^2 Q
|y'|^2 (\ln|y'| - 1) \, . 
\end{eqnarray}
The unexpected and puzzling feature of this solution is that despite
being fully localised, it still breaks more supersymmetries than the
half predicted by the probe analysis \cite{skenderis} or the CFT
approach \cite{dabholkar}. There are several subtle points that might
occur when one looks at D7-brane supergravity solutions, and which
were not taken into account when constructing this solution.  We did
not attempt to fully resolve this apparent mismatch. Instead, we will
comment on possible resolutions in the last section.

\subsection{Non-BPS solutions}

In this section we construct the non-supersymmetric solutions of the
equations (\ref{eqns_s}) and (\ref{eqns_h}).  The classification made
in section~\ref{susies} implies that the D-branes described by our
ansatz, with $(+,-,m,m)$ embeddings (for $m\neq 0$) are
non-supersymmetric. The supergravity analysis agrees with the result
obtained in the appendix, using D-brane probes.

Because of the very simple ansatz that we have, the equations of
motion can be integrated directly, yielding the following solutions
\begin{itemize}
\item D3  $(+,-,1,1)$ :
\begin{equation}
\left\{
\begin{array}{rcl}
H &=& 1 + {Q \over (y^2 + y'^2)^2}  \, ,  \nonumber \\
            S &=& {1\over 32} W^2 (x^2 + x'^2 + y^2 + y'^2  + {Q\over (y^2 + y'^2)}) \, \nonumber \\
\end{array}
\right.
\end{equation}

\item D5  $(+,-,2,2)$  :
\begin{equation}
\left\{
\begin{array}{rcl}
H &=& 1 + {Q \over y^2 + y'^2 }  \, ,  \nonumber \\
           S &=& {1\over 32} W^2 (x^2 + x'^2 + y^2 + y'^2) - {W^2 Q\over 8} \ln (y^2 + y'^2) \, \nonumber \\
\end{array}
\right.
\end{equation}

\item D7  $(+,-,3,3)$  :
\begin{equation}
\label{d7}
\left\{
\begin{array}{rcl}
H &=& 1 + {Q\over 2}  \ln(y^2 + y'^2)   \, ,  \nonumber \\
           S &=& {1\over 32} W^2 (x^2 + x'^2 + y^2 + y'^2) - {2 W^2
           Q\over 32} (y^2 + y'^2) ({1\over 2} \ln (y^2 + y'^2) - 1 )
           \nonumber 
\end{array}
\right.
\end{equation}
\end{itemize}
Let us now analyse the properties of these solutions.  First we see
that, contrary to the supersymmetric D-branes, all solutions are
\emph{fully localised}.  Second, as expected, the D3 and D5-brane
solutions asymptotically reduce to the Hpp-wave~(\ref{KG_10}). The
only case for which this is not true is the D7-brane, which again
exhibits special features. Third, the correction terms to the function
$S$ again blow up for small $y$, $y'$. To see whether this leads to
physical singularities, let us perform an analysis similar to the one
we did in the supersymmetric cases but now on the example of the
non-supersymmetric $D3$-brane. In this case all scalar quantities,
including $R_{\mu\nu} \xi^{\mu} \xi^{\nu}$ (which is again the only
quantity dependent on $S$) are finite. The equations of motion for a
radially infalling particle reduce to
\begin{equation}
\label{pt2}
\dot{r}^2 = {-\epsilon\over \sqrt{1 + {Q\over r^4}}} + 2 p_{+}p_{-} - {1\over
16}W^2 p_{-}^2 r^2 \left(1 + {Q\over r^4}\right) \, ,  
\end{equation} 
where $r^2 = y^2 + y'^2$. If we compare this expression to the
radially infalling particle~(\ref{pt1}) we see that the correction to
the function $S$ (the term proportional to $W^2Q$) now has the opposite
sign. This term becomes dominant for small $r$, causing  a repulsive
behaviour.  While the pure pp-wave causes the focusing
of the geodesics 
towards the $r=0$ geodesics, and the pure $D3$-brane also acts as an
attractor towards the $r=0$ point, the superposed system exhibits a
repulsive behaviour!  Note that this is opposite to the situation we
had for the supersymmetric D3-brane, where the pp-wave was
strengthening the attractive behavior of the source.

Finally, since these D-branes are non-supersymmetric, one should also
raise the issue of the stability. Since these D-branes carry a
conserved charge they for sure cannot decay into the vacuum. Moreover,
a non-supersymmetric D-brane with the same orientation probing this
background does not feel any force. These two properties could point
towards the stability of these embeddings but, clearly, this problem
deserves a separate study and investigation is currently in
progress. In the next section we will propose several possible ways to
address this issue in detail.

\section{Discussion}

To conclude, let us discuss some open questions and problems which
certainly deserve further study.

The first set of problems concerns other supergravity solutions for
D-branes in pp-waves.  The \emph{supersymmetric} D3 and
D5-branes solutions that have 
been constructed in this paper are smeared. As for the cases of
D-brane intersections in Minkowski space, this restriction is
obviously a consequence of the oversimplified form of the ansatz which
has been used. However, we believe that Brinkman coordinates and our
ansatz should serve as a good starting point for the construction of
more complex ansatzes for the fully localised D-brane solutions.

It would be also interesting to find supergravity solutions
corresponding to the families of $1/2$ and $1/4$-BPS D-branes with
nonvanishing worldvolume fluxes that were identified in
\cite{skenderis}. This problem is presumably more complex than the
previous one, since the presence of worldvolume fluxes would lead to
additional nontrivial bulk fluxes at the level of the supergravity
solution.

An apparent puzzle that appeared in this work concerns the \emph{fully
localised, supersymmetric} D7-brane solution~(\ref{d7}). Only in this
case does the number of supersymmetries predicted by the probe
brane/CFT analysis not match with the number of supersymmetries
preserved by our solution.  One speculative resolution of this paradox
is the following.  Since the probe approach does not take into
account the back-reaction of the D-brane on the pp-wave background, some
(super)symmetries might be absent when this effect is incorporated. In
particular we suspect that (``supernumerary'') supersymmetries of the
supergravity solution which in the dual gauge theory are related to
the superconformal symmetries will be absent due to the back-reaction
of the D7-brane on the pp-wave. If this is indeed true then a similar
type of mismatch should already appear between the supergravity
solution for the D7-brane wrapping the $AdS_5\times S^5$ space and the
D-brane probe analysis in the same background.  The probe 
analysis of \cite{skenderis} for this case implies that the number of
supersymmetries is sixteen. The supergravity solution is unfortunately
not fully known \cite{spalinski,afm}. However, the expected dual field
theory is the ${\cal N}=2$, $SU(N)$ gauge theory with $M$
hypermultiplets in the fundamental representation and one
hypermultiplet in the adjoint representation of the gauge group.  If
this theory, whose one-loop beta function is proportional to the
number of $D7$-branes, breaks all superconformal supersymmetries at
the quantum level, then the corresponding supergravity dual does not
possess any supernumerary supersymmetries and preserves only eight
supercharges.
  
To close the discussion of the D7-brane puzzle with a less speculative
comment, let us mention that (super)conformality in the presence of
D7-branes can be restored by canceling their induced RR-charge. A way
to do this is to add orientifold 7-planes. One O7-plane cancels the
charge of 4~D7-branes. In particular, if these charges are locally
canceled, the dilaton is constant and the only effect of the
orientifold is to make a $\mathbb{Z}_2$ identification on the
transverse space of the D7-branes. Therefore, the supergravity
solution for this system in a Hpp-wave is an unmodified Hpp-wave
solution, up to the orbifold identification. Due to this
identification, half of the supersymmetries are projected out, and
hence the supergravity solution preserves one half of the
supersymmetries, which is the same as the (O7/4~D7)-wave-system.

Another set of questions concerns the class of the non-supersymmetric
solutions.  First, all these solutions are fully localised. It
is not clear to us why is this the case. Naively, one would expect
(from the experience one has in flat space) that the simplest
solutions should be those which are characterised by the highest
number of (super)symmetries. Here, we seems to have an interesting
counterexample of this ``intuition''. This, of course, might be a mere
consequence of the very specific coordinates that we were using; there
could be another set of coordinates in which fully localised $1/4$-BPS
solutions would look very simple.  However, we still believe that it
is an observation one should keep in mind when addressing similar
problems of D-branes in other nontrivial spaces.

Second, these non-supersymmetric solutions, as many other solutions
constructed in the literature, display a repulsion type of
behaviour. Here, however for the D3-brane solution, contrary to
standard cases~\cite{kallosh,johnson}, there is no curvature
singularity present behind the repulsion radius.  Except for this
strange behavior it does not seem to exhibit any other unphysical
behaviour.

Third, since the presented solution is non-supersymmetric, it could be
unstable. Notice that the breaking of supersymmetries is due to the
``bad'' orientation of the D-brane in the RR flux of the pp-wave. One
can speculate that possible tachyonic modes would lead to the change
of the orientation of the D-brane in the pp-wave background, and that
the final configuration would be a D-brane with a ``good''
orientation, which preserves some fraction of supersymmetry.  To put
this on a firmer ground, one would have to study these
non-supersymmetric D-branes in more detail, using for instance the
CFT or probe effective action points-of-view to see if their spectra
contain tachyonic open string modes.  If there are no tachyonic modes
present and if these non-BPS D-branes are stable,
then they could have an interesting new role in the context of
\emph{nonsupersymmetric} defect AdS/CFT correspondence. We are
currently investigating this direction.

Finally, let us close this article by saying that we were studying,
from the supergravity point of view, examples of D-brane embeddings
which have been described in the literature using other approaches
\cite{skenderis,dabholkar}. However, it is very likely that there are
many other (non)BPS D-brane embeddings in the Hpp-wave background. A
systematic classification of all these configurations is desirable.

\ 

\

\noindent{\bf Acknowledgements} 

We have benefited from discussions with C.~Bachas, M.~Blau, S.~Kovacs,
C.~Nu\~nez, T.~Ort\'{\i}n, G.~Papadopoulos, K.~Skenderis,
N.~Suryanarayana, M.~Taylor, P.~Townsend and T.~Wiseman. We are
specially grateful to K.~Peeters for many valuble comments and
suggestions.  P.B.  acknowledges PPARC for financial support. This
work is also partially supported by EU contracts HPRN-CT-2000-00122
and HPRN-CT-2000-00148. M.Z. would like to thank CMS for hospitality
during the final stage of this work.

\vfill\eject
\appendix
\section{Some technical details}
\label{app1}
\noindent{\bf Some useful supergravity equations and notations}

\noindent 
The dilaton equation is 
\begin{equation}
\label{eom-dil}
R + 4 (\partial \phi)^2  - 4 \nabla^2 \phi + {1\over 2 \cdot 3!} H^2 = 0 \, .
\end{equation}
The Einstein equations are (here the Ricci scalar has been eliminated
with the use of the dilaton equation)
\begin{equation}
R_{\mu\nu} = 2 \nabla_{\mu} \nabla_{\nu} \phi - {1 \over 4} H_{\mu\xi\rho}H_{nu}^{\xi\rho} + {1\over 4} e^{2\phi} \sum_{n} {(-1)^n \over (n-1)!} T^{(n)}{}_{\mu\nu} \, ,
\end{equation}
where $T^{[n]}{}_{\mu\nu}$ are the energy-momentum  tensors for the RR fields
\begin{equation}
T_{[n]\, \mu\nu} = F_{[n]}{}_{\mu}{}^{\xi_1 \cdots\xi_{(n-1)}} F_{[n]}{}_{\nu\xi_1\cdots \xi_{(n-1)}} - {1\over 2n} g_{\mu\nu} F_{[n]}^{2} \, .
\end{equation}
where $T^{[n]}{}_{\mu\nu}$ are the energy-momentum  tensors for the RR fields
\begin{equation}
T_{[n]\, \mu\nu} = F_{[n]}{}_{\mu}{}^{\xi_1 \cdots\xi_{(n-1)}} F_{[n]}{}_{\nu\xi_1\cdots \xi_{(n-1)}} - {1\over 2n} g_{\mu\nu} F_{[n]}^{2} \, .
\end{equation}
The Bianchi identities and the equations of motion for the RR fields
can be written in compact form as~\cite{pet}
\begin{equation}
\label{BI-RR}
\left\{
\begin{array}{rcl}
dF -H_{[3]}\wedge F & = & 0\, ,\\
& & \\
d H_{[3]} & = & 0\, ,\\
& & \\
dH_{[7]} +\frac{1}{2} {}^{\star} F \wedge F & = & 0\, ,\\
\end{array}
\right.
\end{equation}
\noindent and 
\begin{equation}
\left\{
\begin{array}{rcl}
d{}^{\star}F +H_{[3]}\wedge F & = & 0\, ,\\
& & \\
d \left(e^{-2\phi} {}^{\star} H_{[3]} \right) 
+\frac{1}{2} {}^{\star} F \wedge F & = & 0\, ,\\
& & \\
d\left( e^{2\phi} {}^{\star} H_{[7]}  \right)& = & 0\, ,\\
\end{array}
\right.
\end{equation}
where we are using the notation of~\cite{Douglas,GHT}
in which forms of different degrees are formally combined into a
single entity:
\begin{equation}
\left\{
\begin{array}{rcl}
C & = & C_{[0]} + C_{[1]} + C_{[2]} +\ldots\, , \\
& & \\
F & = & F_{[1]} + F_{[2]} + F_{[3]} +\ldots\, .\\
\end{array}
\right.
\end{equation}
where the dual potentials are defined by the relations
between field strengths
\begin{equation}
\left\{
\begin{array}{rcl}
F_{[10-n]} & = & (-1)^{\left[n/2\right]} {}^{\star}F_{[n]}\, ,  \\
& & \\
H_{[7]} & = & e^{-2\phi} {}^{\star} H_{[3]}\, ,\\
\end{array}
\right.
\end{equation}
 
\ 

\noindent{\bf Vierbein, spin connection and Ricci tensor}

\noindent We choose the following vielbeins
$e^{\underline{m}}={e_{\mu}}^{\underline{m}}dx^{\mu}$ 
\begin{equation}
\begin{array}{ll}
e_{\underline{v}} = e^{\underline{u}} = H^{- {1\over 4}} du \, , & e_{\underline{u}} = e^{\underline{v}} = H^{- {1\over 4}} (dv + S du) \, , \\
- e_{\underline{i}} =  e^{\underline{i}} = H^{-{1\over 4}} dx^i \, , & - e_{\underline{I}} =  e^{\underline{I}} = H^{-{1\over 4}} dx^I \, , \\
 - e_{\underline{a}} = e^{\underline{a}} = H^{1\over 4} dy^{a} \, , & 
 - e_{\underline{A}} = e^{\underline{A}} = H^{1\over 4} dy^{A} \, 
\end{array}
\end{equation}
and the inverse vielbeins 
$\theta_{\underline{m}}={e_{\underline{m}}}^{\mu}\partial_{\mu}$
\begin{eqnarray}
\begin{array}{ll}
\theta_{\underline{u}} = H^{1\over 4} (\partial_u - S \partial_{v})\, , & \theta_{\underline{v}} = H^{1\over 4} \partial_v \, , \\
\theta_{\underline{i}} = H^{1\over 4} \partial_i \, , & \theta_{\underline{I}} = H^{1\over 4} \partial_I \, , \\
\theta_{\underline{a}} = H^{-{1\over 4}} \partial_{a} \, , & 
\theta_{\underline{A}} = H^{-{1\over 4}} \partial_{A} \, 
\end{array}
\end{eqnarray}
and we use flat light cone metric $\eta^{\underline{uv}}= 1$, $\eta^{\underline{ij}}= -1$.

The corresponding spin connection is 
\begin{equation}
\begin{array}{ll}
\omega_{\underline{uui}} = - H^{1\over 4} \partial_{i} S \, ,  & \omega_{\underline{uuI}} = - H^{1\over 4} \partial_{I} S \, , \\
\omega_{\underline{u}\underline{u}\underline{a}} = - H^{-{1\over 4}}\partial_{a} S \, & \omega_{\underline{uuA}} = - H^{-{1\over 4}}\partial_{A} S 
\\
\omega_{\underline{uva}} = {1\over 4} H^{- {5\over 4}} \partial_{a} H \, ,  & \omega_{\underline{uvA}} = {1\over 4} H^{- {5\over 4}}\partial_{A} H \, \\
\omega_{\underline{iaj}} = H^{-{5\over 4}} \partial_{a} H \delta_{ij} \, ,  & \omega_{\underline{IaJ}} = H^{-{5\over 4}} \partial_{a} H \delta_{IJ} \, , \\
\omega_{\underline{iAj}} = H^{-{5\over 4}} \partial_{A} H \delta_{ij} \, ,  & \omega_{\underline{IAJ}} = H^{-{5\over 4}} \partial_{A} H \delta_{IJ} \, , \\
\omega_{\underline{abc}} = {1\over 2} H^{-{5\over
4}}\eta_{a[b}\partial_{c]} H \, , & \omega_{\underline{abC}} = {1\over 4} H^{-{5\over 4}}\eta_{ab} \partial_{C} H \, , \\
\omega_{\underline{ABc}} = {1\over 4} H^{-{5\over
4}}\eta_{AB}\partial_{c} H \, ,  & \omega_{\underline{ABC}} = {1\over 2} H^{-{5\over 4}}\eta_{A[B} \partial_{C]} H \, .

\end{array}
\end{equation}
Finally, the components of the Ricci tensor for the metric~(\ref{ansatza}) are given by 
\begin{eqnarray}
R_{vv} &=& 0 \, , \nonumber \\
R_{uu} &=& \partial_{i} \partial^i S + H^{-1} \partial_{\hat{a}} \partial^{\hat{a}} S + {(3-p)\over 2} H^{-1} \partial_{\hat{a}}\partial^{\hat{a}} \log H \, , \nonumber \\
&-& {1\over 2} S ( H^{-2} \partial_a \partial^a H - {p-1 \over 2} H^{-1} \partial_{a} \log H \partial^a \log H ) \,  , \nonumber \\
R_{uv} &=& - {1\over 4} (H^{-2} \partial_{a} \partial^{a} H - {p-1 \over 2} H^{-1} \partial_{a} \log H \partial^{a} \log H) \, , \nonumber \\
R_{ij} &=& - {1\over 4} \eta_{ij} (H^{-2} \partial_{a} \partial^{a} H - {p-1 \over 2} H^{-1} \partial_{a} \log H \partial^{a} \log H) \, , \nonumber \\
R_{ab} &=& {1\over 4} \eta_{ab} H^{-1} \partial_{c} \partial^{c} H -
{p-3 \over 2} H^{-1} \partial_a \partial_b H - {p-1 \over 8} \eta_{ab}
\partial_{c} \log H \partial^{c} \log H \nonumber\\
 &+& {3p - 7 \over 4} \partial_{a} \log H \partial_{b} \log H \, . \nonumber 
\end{eqnarray}

\section{Probe analysis}
\label{app2}
In the this appendix we repeat some of the D-brane probe analysis of
\cite{skenderis}\footnote{The original discussion of \cite{skenderis}
contained a mistake in counting the number of supersymmetries
preserved by the D-branes with the embedding $(+,-,m,m)$ for
$(m=1,2,3)$. This was corrected in the revised version. We are
grateful to the authors of \cite{skenderis} for correspondence
regarding this issue.}, in order to make contact with the
supersymmetry analysis from the supergravity point of view as
presented in the main text.

For negative chirality type IIB spinors,
\begin{equation}
\Gamma^{(11)} \epsilon = - \epsilon, 
\label{chirality}
\end{equation}
the Killing spinors of the Hpp-wave are given by \cite{blau}
\begin{eqnarray}
\epsilon(u, z) &=& \left(1 - \frac{i}{2} \sum_{\mu=1}^4 \gamma_{\underline{v}}
(z^\mu \gamma_{\underline{\mu}}
I + z^{\mu+4} \gamma_{\underline{\mu+4}} J) \right) \nonumber\\
&&~~~~~~ \left(\cos \frac{1}{2}u - i I \sin\frac{1}{2}u \right) \left(\cos\frac{1}{2}u - i J \sin\frac{1}{2}u \right)
\chi
\label{killing}
\end{eqnarray}
where we have defined $I=\gamma_{\underline{1234}}$,
$J=\gamma_{\underline{5678}}$. The  matrices $\gamma_{\underline{u}}$,
$\gamma_{\underline{v}}$ and 
$\gamma_{\underline{\mu}}$ ($\mu=1, \ldots, 8$) are tangent space,
$32 \times 32$ 
Dirac matrices and the complex spinor $\chi$, subject to the chirality
projection (\ref{chirality}), can be decomposed as $\chi=\lambda + i
\eta$, where $\lambda$ and $\eta$ are constant, real, negative
chirality spinors.

In the absence of worldvolume fluxes, the
kappa-symmetry projection operator, for a D-brane with the 
$(+,-,m,n)$ embedding, is given by
\begin{equation}
\Gamma = \gamma_{\underline{uva_1\ldots a_mb_1\ldots b_n}} K^{p+1 \over 2} I \equiv Q
K^{p+1\over 2}I \, , 
\end{equation}
where $K$ acts by complex conjugation and $I$ by multiplication by $-i$.

The amount of supersymmetry preserved by the D-brane probe can be
determined from the condition
\begin{equation}
\label{gamma}
\Gamma \epsilon = \epsilon \, ,
\end{equation}
where $\epsilon$ is the Killing spinor of the background in the which
probe is embedded, and it is evaluated on the worldvolume of the
D-brane.  For the $D3$-brane, the condition (\ref{gamma}) reduces to
\begin{equation} 
\epsilon = -i Q \epsilon \, , 
\label{kappa1}
\end{equation}
while for the $D1$ and $D5$-branes, it reduces to 
\begin{equation} 
\epsilon = i Q \epsilon^* \, .
\label{kappa2}
\end{equation}
Let use consider the D3-brane in detail for the two possible
embeddings, $(+,-,2,0)$ and $(+,-,1,1)$.  In the first case, the
projection operator is $Q_1 = \gamma_{\underline{uv12}}$ while in the
second case $Q_2=\gamma_{\underline{uv15}}$. A crucial property to
note is that $Q_1$ commutes with $I$ and $J$ while $Q_2$ anticommutes
with $I$ and $J$.

Plugging the Killing spinor (\ref{killing}) into the equation
(\ref{kappa1}) and imposing that the equation should hold for any
$u$ lead to the following three $z$-dependent equations
\begin{eqnarray}
&&\left(1- \frac{i}{2} \sum_{a=1}^{4} \gamma_{\underline{v}} 
(z^\mu \gamma_{\underline{\mu}}
I + z^{\mu+4} \gamma_{\underline{\mu+4}} J) \right) \chi = \nonumber\\
&&~~~~~~~~~~~~~~~-i Q 
\left(1-\frac{i}{2}\sum_{a=1}^4 \gamma_{\underline{v}} 
(z^\mu \gamma_{\underline{\mu}}
I + z^{\mu+4} \gamma_{\underline{\mu+4}} J)  \right)
\chi \nonumber\\ 
&&\left(1-\frac{i}{2} \sum_{a=1}^4 \gamma_{\underline{v}} 
(z^\mu \gamma_{\underline{\mu}}
I + z^{\mu+4} \gamma_{\underline{\mu+4}} J)\right)(I+J) \chi = \nonumber\\
&&~~~~~~~~~~~~~~~-i Q 
\left(1-\frac{i}{2}
\sum_{a=1}^4 \gamma_{\underline{v}} (z^\mu \gamma_{\underline{\mu}}
I + z^{\mu+4} \gamma_{\underline{\mu+4}} J) \right)(I+J)
\chi \nonumber\\ 
&&\left(1-\frac{i}{2} \sum_{a=1}^4 \gamma_{\underline{v}} (z^\mu
\gamma_{\underline{\mu}} 
I + z^{\mu+4} \gamma_{\underline{\mu+4}} J) \right) IJ \chi = \nonumber\\
&&~~~~~~~~~~~~~~~-i Q 
\left(1-\frac{i}{2}
\sum_{a=1}^4 \gamma_{\underline{v}} (z^\mu
\gamma_{\underline{\mu}} 
I + z^{\mu+4} \gamma_{\underline{\mu+4}} J)\right) IJ  
\chi 
\label{eqn1}
\end{eqnarray}
It is easy to see that, due to the identity $(IJ)^2=1$, the first and
the last equations are identical. 

Next, we use the fact that the previous equations have to be satisfied
for any value of the coordinates $x^{\hat{\imath}}$ along the D-brane
worldvolume. Here, we will consider only D-branes sitting at the
origin of the space in the Brinkman coordinates, and we choose the
static gauge for the worldvolume directions. Therefore, all the
coordinates transverse to the D-brane in~(\ref{eqn1}) can be set to
zero and we only keep the dependence of the spinor on the worldvolume
coordinates.

For the $(+,-,2,0)$ embedding, we obtain the system of equations
\begin{eqnarray}
\label{c1}
&&\chi = -i Q_1 \chi \\
\label{c2}
&&(I+J)\chi = -i Q_1 (I+J) \chi \\
\label{c3}
&&\gamma_{\underline{v}} \gamma_{\underline{a}} I \chi = -i Q_1
\gamma_{\underline{v}} \gamma_{\underline{a}} I 
\chi \\
\label{c4}
&&\gamma_{\underline{v}} \gamma_{\underline{a}} I (I+J) \chi = -i Q_1 \gamma_{\underline{v}}
\gamma_{\underline{a}} I (I+J)  
\chi \, ,
\end{eqnarray}
where $a=1,2$ indices correspond to directions $x^1,x^2$ along which
the D-brane extends. The last two equations come from the worldvolume
dependent part of the equation~(\ref{eqn1}).

Since for this embedding $I$ and $J$ commute with $Q_1$, equation
(\ref{c2}) is by virtue of equation (\ref{c1}) automatically
satisfied. Similarly, using (\ref{c3}), equation (\ref{c4}) is also
satisfied. Finally, since $Q_1$ anticommutes with both
$\gamma_{\underline{v}}$ and $\gamma_{\underline{i}}$ it is easy to
see that if (\ref{c1}) holds, equation (\ref{c3}) is also
satisfied. Hence, the only independent condition is obtained from the
first equation~(\ref{c1}). The real and imaginary parts of this
equation both lead to the following constraint on the spinors
\begin{equation}
\lambda = Q \eta \, .
\end{equation} 
Hence, the $(+,-,2,0)$ D3-brane sitting at the origin of the space 
breaks exactly one half of the supersymmetries of the Hpp-wave background.  

Let us now consider the $(+,-,1,1)$ embedding of the D3-brane. The
kappa-symmetry projection leads to the following set of equations
\begin{eqnarray}
\label{cc1}
&&\chi = -i Q_2 \chi \\
\label{cc2}
&&(I+J)\chi = -i Q_2 (I+J) \chi \\
\label{cc3}
&&\gamma_{\underline{v}} \gamma_{\underline{a}} I \chi = -i Q_2
\gamma_{\underline{v}} \gamma_{\underline{a}} I 
\chi \\
\label{cc4}
&&\gamma_{\underline{v}} \gamma_{\underline{a}} I (I+J) \chi = -i Q_2 \gamma_{\underline{v}}
\gamma_{\underline{a}} I (I+J) 
\chi \, , 
\end{eqnarray}
where now $a=1, 5$. The last two sets of equations come from the
worldvolume coordinates ($x^1 = z^1, x^{\prime 1}=z^5$) dependent part
of the relations~(\ref{eqn1}).  
First, since $I$ and $J$ now anticommute with $Q_2$, we see that the
first and second equations are compatible only if we impose that 
\begin{equation}
(I+J)\chi = 0.
\label{proj1}
\end{equation}
This constraint implies that the equations~(\ref{cc4}) of the above
system are trivially satisfied. Let us consider the third
equation~(\ref{cc3}). Since $I$, $\gamma_{\underline{v}}$ and
$\gamma_{\underline{a}}$ all anticommute with $Q_2$, after 
multiplication of the equation (\ref{cc3}) with the (invertible) matrix
$\gamma_{\underline{a}} I$, one obtains the condition
\begin{equation}
\gamma_{\underline{v}} \chi = i \gamma_{\underline{v}} Q_2 \chi \, ,
\label{constr1}
\end{equation}
which, when combined with (\ref{cc1}), leads to 
\begin{equation}
\gamma_{\underline{v}} \chi = 0.
\label{proj2}
\end{equation}
However, this condition implies, together with the negative chirality
condition for spinors (\ref{chirality}) and the constraint
(\ref{proj1}), that $\chi = 0$, as follows. By multiplying
(\ref{proj1}) with the invertible matrix $\gamma_{\underline{uv}} I$,
after the use of (\ref{chirality}), one gets the condition
\begin{equation}
\gamma_{\underline{uv}} \chi = \chi \, 
\end{equation} 
which is compatible with (\ref{proj2}) only if $\chi=0$. Therefore, no
supersymmetry is preserved by the D3-brane with embedding
$(+,-,1,1)$. A similar analysis for the $(+,-,2,2)$ D5-brane or the
$(+,-,3,3)$ D7-branes leads to the same conclusions.

The D-string is different from higher dimensional D$p$-branes since
in this case there are no worldvolume directions transverse to the
pp-wave. This means that the constraint~(\ref{proj2}) associated to
these worldvolume coordinates is absent, and one only has to impose
the conditions~(\ref{cc1}) and~(\ref{proj1}). These two conditions can
be satisfied simultaneously leading to the conclusion that the
D-string is $1/4$-BPS.

\end{document}